\documentclass[12pt,showpacs,preprintnumbers,amsmath,amssymb]{revtex4}
\usepackage[english]{babel}
\usepackage[latin2]{inputenc}
\usepackage{epsfig}

\begin{document}

\title{Correlated Doping in Semiconductors:\\ The Role of
Donors in III-V Diluted Magnetic Semiconductors}

\author{J.~Ma\v{s}ek$^{a}$, I.~Turek$^{b}$, V.~Drchal$^{a}$,
J.~Kudrnovsk\'{y}$^{a}$, and F.~M\'{a}ca$^{a}$}

\affiliation{$^{a}$Institute of Physics, AS CR, Na Slovance 2,
CZ-182 21 Prague 8 \\ $^{b}$Institute of Physics of Materials, AS
CR, \v{Z}i\v{z}kova 22, CZ-616 62 Brno\\ and Faculty of
Mathematics and Physics, Charles University, Prague}


\begin{abstract}
We investigate the compositional dependence of the total energy of
the mixed crystals (Ga,Mn)As co--doped with As, Sn, and Zn. Using
the {\sl ab initio} LMTO--CPA method we find a correlation between
the incorporation of acceptors (Mn, Zn) and donors (Sn, antisite
As). In particular, the formation energy of As$_{\rm Ga}$ is
reduced by approx. 0.1 eV in the presence of Mn, and vice versa.
This leads to the self--compensating behavior of (Ga,Mn)As.
\end{abstract}


\maketitle

\section{Introduction}

The Mn doping of the III-V semiconductors has two principal
effects in the III-V diluted magnetic semiconductors (DMS). The
correlated d-electrons at Mn atoms form local magnetic moments and
the hybridization of the d-states with the band states results in
various magnetoelectric and magnetooptical phenomena \cite{Ohno99}
. In addition, Mn atoms substituted for a trivalent cation act as
acceptors and introduce holes into the valence band. It is now
generally accepted \cite{Dietl00} that the ferromagnetic coupling
between the local moments in the III-V DMS is mediated by the
mobile holes in the valence band.

In reality, however, the number of the holes is much smaller than
the nominal concentration of Mn \cite{Beschoten99, Edmonds02} This
indicates that a large amount of compensating donors is present,
with a strong effect on both conductivity and Curie temperature of
these materials. As pointed out in \cite{Maca02}, the nearly
constant (or even decreasing) doping efficiency of order $0.1 \sim
0.2$ can be explained only assuming that the number of the donors
increases proportionally to the concentration of Mn. This was the
reason to suggest that Mn atoms in the interstitial positions -
being double donors - might have a principal role in the
compensation \cite{Masek01}.

Another favorite candidate for the compensating donor is an As
antisite defect. These defects are well known in crystals grown
with an excess As, in particular in the (Ga,Mn)As films
\cite{Shimizu99}. So far, however, there was no reason to expect
any correlation between the concentration of As antisites and the
concentration of Mn.

Only recently, an increase of the number of the As antisites was
indicated by comparing Curie temperature obtained from ab initio
calculations for (Ga,Mn,As)As mixed crystals \cite{Kudrnov02} with
experimental data. This finding opens a question whether and why
the number of the As antisites is correlated with the level of Mn
doping.

We investigate the correlation between the donors and acceptors in
partly covalent III-V semiconductors such as GaAs. The cohesion
energy of the covalent networks has a maximum if the Fermi energy
lies within a band gap (Ioffe--Regel's rule \cite{Ioffe}) .
Whenever the Fermi energy is situated in the valence or conduction
band the strength of the bonds is reduced because the unfilled
bonding states or occupied antibonding states appear,
respectively. It is natural to expect that this mechanism,
connected with the changes of the Fermi energy position in
dependence on the compensation, leads to some kind of acceptor -
donor correlation.

\section{Correlation energy}

We consider a crystal doped with both acceptors and donors. For
simplicity, we assume that both acceptors and donors are
substitutional impurities, which is the case of both As antisites
and Mn in GaAs. Their concentrations are $x_{A}$ and $x_{D}$,
respectively. The total energy of the doped crystal, normalized to
a unit cell, is $W(x_{A},x_{D})$. We show first that the
derivatives of $W(x_{A},x_{D})$ with respect to the concentrations
$x_{A}$ and $x_{D}$ determine the formation energies of the
defects, and also the correlation energy of the co-doping.

We start with a large unit cell (LUC) consisting of $N$ unit cells
of the mixed crystal. The formation energy $E^{A}$ of an acceptor
A is defined as the reaction energy of the substitution process

\begin{center}
 LUC$^{(X)}$ + A  $\longrightarrow$ LUC$^{(A)}$ + X.
\end{center}

Here, LUC$^{(A)}$  is a large unit cell with one extra acceptor A
replacing an atom X of the original LUC$^{(X)}$. In our notation,
the corresponding reaction energy is
\begin{equation}
E^{A}(x_{A},x_{D}) = N \cdot ( W(x_{A} + 1/N,x_{D}) -
W(x_{A},x_{D})) + E_{atom}(X) - E_{atom}(A).
\end{equation}
The last two terms in Eq.~(1) are the total energies of
free-standing atoms X and A, respectively. The additional constant
$E_{atom}(X) - E_{atom}(A)$, though crucial for the correct
absolute value of the formation energy, does not depend on the
actual composition of the material and so it is not important for
the concentration--dependent effects we have in mind. With
increasing size of the large unit cell, $N \rightarrow \infty$,
the first term in Eq. (1) approaches the derivative of
$W(x_{A},x_{D})$ with respect to $x_{A}$. We have
\begin{equation}
E^{A}(x_{A},x_{D}) = \frac{\partial W(x_{A},x_{D})}{\partial
x_{A}} + E_{atom}(X) - E_{atom}(A),
\end{equation}
in a close analogy with the definition of the chemical potential.
Similarly, the formation energy of a donor D substituting for an
atom Y is
\begin{equation}
E^{D}(x_{A},x_{D}) = \frac{\partial W(x_{A},x_{D})}{\partial
x_{D}} + E_{atom}(Y) - E_{atom}(D),
\end{equation}
Finally, the compositional dependence of the formation energies
can be characterized by the correlation energy
\begin{equation}
K(x_{A},x_{D}) = \frac{\partial E^{A}}{\partial x_{D}} =
\frac{\partial E^{D}}{dx_{A}} = \frac{\partial
^{2}W(x_{A},x_{D})}{\partial x_{A} \partial x_{D}}.
\end{equation}
The correlation energy can be alternatively, for finite large unit
cells, expressed in terms of the four total energies corresponding
to the reference LUC$^{(XY)}$ and to related systems with an extra
acceptor (LUC$^{(AY)}$), extra donor (LUC$^{(XD)}$), and both
acceptor and donor (LUC$^{(AD)}$),
\begin{equation}
K = W\{LUC^{(AD)}\} - W\{LUC^{(AY)}\} - W\{LUC^{(XD)}\} +
W\{LUC^{(XY)}\}
\end{equation}
The correlation energy $K(x_{A},x_{D})$ is positive if the
formation energy of one impurity increases in the presence of the
other. This means, in the case of semiconductors, that the
material tends to be either n-type or p-type rather than a
compensated semiconductor. On the other hand, negative correlation
energy indicates that the presence of impurities of one kind makes
the incorporation of the other dopants easier. In this case we can
speak about a preferential compensation.

\begin{figure}[tbp]
\begin{center}
\epsfig{file=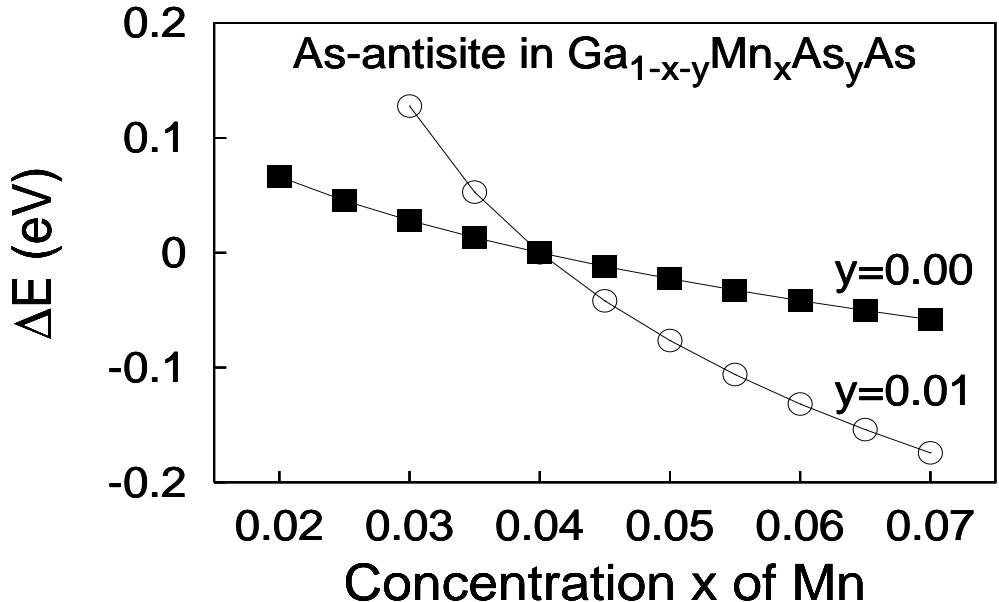, width=84mm, height=146mm, angle=270}
\end{center}
\caption{Dependence of the As antisite formation energy on the
concentration $x$ of Mn acceptors for two series,
Ga$_{1-x}$Mn$_{x}$As (boxes, $y=0.00$) and
Ga$_{0.99-x}$Mn$_{x}$As$_{0.01}$As (circles, $y=0.01$). The
changes $\Delta E$ of the formation energy with respect to the
reference systems with $x = 0.04$ are shown.}
\end{figure}

To investigate the correlation among the dopants, we use the
coherent potential approximation (CPA) \cite{CPA}. There are two
reasons for this. First of all, CPA describes the
configurationally averaged behavior of the mixed systems rather
than particular arrangements of the impurities. This fits well to
the thermodynamical Yoffe-Regel mechanism. In addition, the CPA is
applicable to impure crystals with arbitrary concentrations of the
dopants and it is particularly suitable for the description of the
compositional dependence of the electronic properties.

\section{Results}

We considered a series of multicomponent mixed crystals
Ga$_{1-x-y}$A$_{x}$D$_{y}$As with acceptors A=Mn, Zn and donors
D=As, Sn. The highly degenerate nonmagnetic mixed crystals with Zn
were studied in parallel to the diluted magnetic semiconductor
(Ga,Mn)As to make clear that the acceptor--donor correlation does
not depend on the presence of the magnetic moments. With our
choice, all substitutions take place in the cationic sublattice.
Only doping giving p--type materials was considered.

The {\sl ab initio} TB-LMTO version of the CPA \cite{LMTOCPA} was
used to calculate the electronic structure of the impure crystals
and of their total energies. To obtain the derivatives of the
total energy with respect to the concentrations, we varied both
$x$ and $y$ on a fine mesh with $\delta x, \delta y =0.005$.

Table I summarizes the calculated correlation energies $K(x,0)$.
The correlation energy of the co--doping is negative in all
considered materials. The absolute value of $K$ is of order of a
few electronvolts and it generally decreases with increasing level
of the doping. The weakest correlation if found in (Ga,Mn)As
compensated with Sn. The correlation energies for Sn are
approximately two times smaller than the correlation energies for
the As antisite, which is a double donor, as expected.\\

\noindent TABLE~~I\\ Correlation energy for acceptors (A) and
donors (D) in p--type mixed crystals Ga$_{1-x}$A$_{x}$As at
various levels of p--type doping.
\begin{center}
\begin{tabular}{|c|c|c|c|}
\hline
 {\it x} & A=Mn, D=As & A=Zn, D=As & A=Mn, D=Sn \\ \hline
 0.04    & -3.36 eV   & -5.48 eV   & -1.39 eV   \\
 0.05    & -2.67 eV   & -4.67 eV   & -1.22 eV   \\
 0.06    & -2.21 eV   & -4.19 eV   & -1.09 eV   \\
 \hline
 \end{tabular}
\\~\\
\end{center}

The negative correlation energy of the co-doping means that the
formation energies of both acceptors and donors decrease in the
presence of the compensating impurities. This is shown explicitely
for the substitutional Mn and for the As antisite defect in Figs.
1 and 2. To avoid the technical problem with the additional
constants in Eqs. (2,3), we do not plot the entire formation
energies ($\approx$ 2 eV), but only their changes with respect to
the reference material, mostly Ga$_{0.96}$Mn$_{0.04}$As.

Fig. 1 shows how the formation energy of the As antisite defect
changes with the concentration of Mn. In addition to the main
series of the data for Ga$_{1-x}$Mn$_{x}$As we considered also a
series of mixed crystals Ga$_{0.99-x}$Mn$_{x}$As$_{0.01}$As
already containing a small portion of the As antisites. In both
cases, formation energy is reduced by approx. 0.1 eV if the Mn
concentration increases by a few atomic percent.

\begin{figure}[tbp]
\begin{center}
\epsfig{file=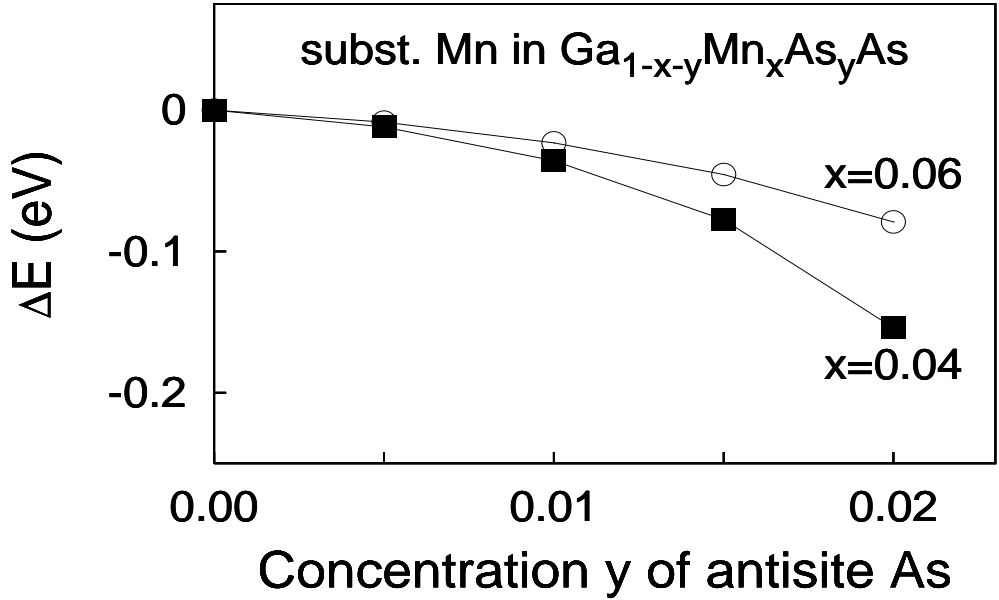, width=84mm, height=146mm, angle=270}
\end{center}
\caption{Change of formation energy, $\Delta E$, of substitutional
Mn in Ga$_{0.96-y}$Mn$_{0.04}$As$_{y}$As (boxes) and
Ga$_{0.94-y}$Mn$_{0.06}$As$_{y}$As (circles) due to the As
antisite defects as a function of their concentration~$y$.}
\end{figure}

The formation energy of the substitutional Mn was calculated for
two series of mixed crystals, Ga$_{0.96-y}$Mn$_{0.04}$As$_{y}$As
and Ga$_{0.94-y}$Mn$_{0.06}$As$_{y}$As. Its variation with
increasing ammount of the As antisites is shown in Fig. 2. The
changes of the formation energy are again of order of 0.1 eV.

To explain the results we turn to a simple model. In the
one-electron picture, the total energy can be expressed in terms
of the density of states $g(E)$,
\begin{equation}
W = \int_{-\infty}^{E_{F}} g(E)\cdot (E_{F} - E) dE.
\end{equation}
The dependence of $W(x_{A},x_{D})$ on the chemical composition
arises both from the redistribution of the electron states in the
valence band due to the impurities and from the changes of the
position of the Fermi level. In the case that $g(E_{F})$ does not
change much with the chemical composition, the total energy
depends on the concentrations of the impurities mostly via the
position of the Fermi level. Because the extra acceptors (donors)
push the Fermi level to lower (higher) energies, the correlation
energy for the acceptor-donor co--doping is expected, according to
Eqs. (5) and (6), to be always negative. Morever, as the
variations of the $E_{F}$ are inversely proportional to
$g(E_{F})$, one can expect that also $W \cdot g(E_{F}) \approx
const.$. This is in a good agreement with the calculated decrease
of the correlation energy with increasing concentration of Mn (cf.
Table~I).

\section{Summary}

We showed that the formation energy of As antisite defects in GaAs
decreases with increasing concentration of Mn or Zn in the
cationic sublattice. The formation energy is reduced by approx.
0.1 eV so that the number of these native defects can be largely
enhanced in the presence of Mn. This effect may contribute to the
self-compensation behavior of (Ga,Mn)As mixed crystals.

At the same time, the formation energy of the substitutional Mn is
similarly reduced in the presence of the As antisites or by
intentional co--doping with Sn. This can be interpreted that the
presence of the donors, either native defects or intentional
dopants, is important for an improved solubility of Mn in III-V
materials.

Finally, the correlation between donors and acceptors seems to be
a general feature tending to a selfcompensation.\\

\begin{acknowledgments}
The financial support was provided by the Academy of Sciences of
the Czech Republic (Grant No. A1010214), by the Grant Agency of
the Czech Republic (202/00/0122), and by RTN project
"Computational Magnetoelectronics" of the European Commission
(HPRN-CT-2000-00143).
\end{acknowledgments}


\end{document}